# A historical introduction to the symmetries of magnetic structures:

Part 1. Early quantum theory, neutron powder diffraction and the colored space groups

Andrew S. Wills

Department of Chemistry, University College London, 20 Gordon Street, London, WC1H 0AJ.

a.s.wills@ucl.ac.uk

Abstract

This paper introduces the historical development of the symmetries for describing magnetic structures culminating in the derivation of the black and white and coloured space groups. Beginning from Langevin's model of the Curie law, it aims to show the challenges that magnetic ordering presented and how different symmetry frameworks were developed to meet them. As well as explaining core ideas, later papers will show how the different schemes are connected. With these goals in mind, the maths related is kept to the minimum required for clarity. Those wishing to learn more details are invited to engage with the references.

As well as looking back and reviewing the development of magnetic symmetry over time, particular attention is spent on explaining where the concept of *time-reversal* has been applied. That time-reversal has different meaning in classical and quantum mechanical situations has created confusions that continue to propagate.

1 Presenting the challenge

In October 1949, Clifford Shull and Samuel Smart published the first magnetic neutron diffraction pattern (1949), Figure 1. Their experimental observation that the magnetic structure of MnO had a unit cell that was twice that of the nuclear along each of the crystallographic axis, was conclusive proof that the predictions of antiferromagnetism by Néel were correct. It also laid a very clear challenge to crystallography as the conventional Fedorov and Schoenflies space groups of crystallography (Hahn, 1996) were insufficient to characterise the alternation in direction of neighbouring moments. The development that followed needed to expand and generalise the concept of symmetry, and still remains a source of argument in the scientific literature.

2 Early quantum mechanics - from paramagnets to ferromagnets

One of the key steps in the development of magnetism at the atomic level, was Paul Langevin's theoretical model of the Curie Law, the empirical finding by Pierre Curie (Curie 1895) that many materials show an inverse relationship between the susceptibility and temperature (Figure 2),

$$\chi \propto \frac{1}{T} \qquad (1)$$

Langevin explained this observation using a model of Ampère's currents in which electrons move in intra-atomic or intra-molecular closed orbits. Orientation of the orbits towards an external magnetic field that would increase the magnetisation is opposed by thermal randomisation, following the Boltzmann distribution. The temperature-dependent compromise was paramagnetism (Langevin, 1905).

Pierre Weiss quickly built upon Langevin's model, which was based in independent electronic orbits (magnetic moments), by assuming that the atomic moments could interact with each other. He expressed this interaction through a local magnetic field which he called the 'local molecular field', $\boldsymbol{H}_m$ that was proportional to the magnetisation, $\boldsymbol{J}$, according to

$$\boldsymbol{H}_m = n\boldsymbol{J} \qquad (2)$$

where $n$ is a constant (Weiss, 1906; Weiss, 1907). This led to a modification of the Curie law that is now known as the Curie-Weiss law, $\chi \propto \frac{C}{T-nC}$. When $n$ is positive, the magnetic moments would tend to align parallel with $\boldsymbol{H}_m$, causing the inverse susceptibility $\frac{1}{\chi}$ to decrease linearly with temperature until below an ordering temperature $nC$, the internal field is large enough for the material to become spontaneously magnetised. The ordering temperature $nC$, which is also known as the Weiss temperature is now commonly given the symbol, $\theta_W$. The value of the inverse susceptibility is then reduced from that of the Curie law according to

$$\frac{1}{\chi'} = \frac{1}{\chi} - n \qquad (3)$$

where $\frac{1}{\chi'}$ corresponds to the ferromagnetic term, and $\frac{1}{\chi}$ the paramagnetic term.

Weiss's model had created a natural explanation for ferromagnetism. It also made apparent a surprisingly large strength of the local molecular field in several insulators: the values of $\boldsymbol{H_m}$, were far too great to be explained using direct short-ranged interactions between magnetic dipoles. (This will be returned to in Section 4.)

3 The case of negative $n$

As well as explaining ferromagnetism, Weiss's model opened up the question of what the case with a negative $n$ would correspond to. At this point, the answer was far from obvious as the only known magnetic states were those of diamagnetism, paramagnetism and ferromagnetism. A notable solution to this problem came some 25 years later, in 1932 from the Ph.D. work of a young theorist that worked as an assistant to Weiss in Strasbourg, Louis Néel (1932).

Following closely the phenomenological local molecular field picture that Weiss had developed, Néel knew from the Curie-Weiss law that at $T = 0$, $\chi$ would to take a finite value equal to $-1/n$ (Figure 3). He proposed that the case of negative $n$ would relate to a tendency of the spins on neighbouring magnetic moments to anti-align. Later in 1936, he would expand upon this picture by proposing that the low temperature magnetic order was made up of two sublattices, *A* and *B*, each of which consisted of parallel spins. *A* and *B* were magnetised in opposite directions so that their net magnetisation was zero at low temperature (Néel, 1936). A moment of sublattice *A* would then be subject to two molecular fields, $n_{aa}\boldsymbol{J_a}$ and $n_{ab}\boldsymbol{J_b}$, proportional to the magnetisations $\boldsymbol{J_a}$ and $\boldsymbol{J_b}$. For the current situation $n_{ab}$ is negative, as positive would correspond to ferromagnetism:

$$\boldsymbol{H_a} = n_{aa}\boldsymbol{J_a} + n_{ab}\boldsymbol{J_b} \qquad (4)$$

The following year, the American Francis Bitter named this situation 'antiferromagnetism' as part of his development of a quantum mechanical treatment of ferromagnetic and antiferromagnetic ordering (1938).

Néel and Bitter were not the only ones interested in the situation of a negative Weiss field. Independently, and around the same time, the Russian theorist Lev Landau, who would become one of the great condensed matter theorists, explored a theoretical model based on the crystal structures of various transition metal halides, in which slabs of ferromagnetically coupled spins were coupled together through a negative Weiss field (antiferromagnetism) (Landau, 1933). He would develop further the style of workings used for this model, an expansion of the terms in the free energy, into a framework that would later become the backbone of condensed matter physics — the celebrated Landau model for a continuous phase transition.

As Landau explored the concepts of antiferromagnetism, he became increasingly convinced that the related ground states were in fact not eigenstates of the Heisenberg Hamiltonian as they were unstable with respect to quantum fluctuations.

Concerns over the stability of Néel's picture of antiferromagnetism, and the antiparallel ground state, culminated at the first *International Conference in Magnetism* in Strasbourg during May of 1939 organised by Weiss. The note on the meeting published in 1946 showed how the researchers had a clear view of how history would soon intervene in magnetism research (Barnett, 1946): the meeting was restricted to 18 reports that were circulated in advance and the gathering was focused around six discussion sessions lasting three hours each. At this meeting Néel was unable to defend his phenomenological model of antiferromagnetism against the perceived requirements of quantum mechanics, and the tone towards it finished as one of doubt.

By the end of 1940, Weiss had died, his world-leading laboratories for magnetism had been plundered, and its workers scattered or deported to Germany for forced labour. Many of the leading names in fundamental magnetism research moved to more applied areas as demanded by the war efforts.

4 Magnetic neutron scattering - experimental validation

It would be in America that antiferromagnetism would finally gain a firm theoretical foundation following the work of John Van Vleck (1941), in a *tour de force* that rivalled that which he had already accomplished to explain paramagnetism. Though, this is not to say that science had a clear picture of what antiferromagnetism would involve. This would become tangible following a legacy of the war that would have a positive effect on fundamental magnetism.

As the neutron sources created for the Manhattan project were finding new uses for fundamental science, Ernest Wollan led a series of fundamental investigations into neutron diffraction by monochromatised neutrons at the Oak Ridge National Laboratory (Shull *et al*, 1951, Mason *et al.*, 2013). Part of this, a study by Clifford Shull and Samuel Smart of MnO (Shull and Smart, 1949), generated some very surprising results (Anderson, 2011) — they saw *new diffraction peaks* at low temperature (Figure 1). Comparison of data taken above and below the antiferromagnetic transition showed these peaks came from magnetic scattering and an ordered structure with lattice parameters that were twice those of the nuclear crystal structure.

Following a model of neutron scattering from atomic magnetic moments developed by Halpern and Johnson (Halpern and Johnson, 1939), they modelled the data using a structure (Figure 5) in which neighbouring $Mn^{2+}$ spins formed sublattices that were antiparallel to each other (Shull *et al*, 1951), exactly as predicted by Néel. His simple phenomenological model was shown to be correct and quantum mechanics had been shown to be less restricted than Landau had thought — within its possibilities was the flexibility to support antiferromagnetic ground states.

The clear image of an antiferromagnetic ground state would also re-cast the question concerning the surprising strength of the magnetic interactions in insulators, where the magnetic ions could not interact directly as they were separated by intervening ions. The theorist Philip Anderson responded in 1951 with a simplification of Kramer's model of superexchange that focussed on the spin-dependent part of the Hamiltonian, and made accessible a linkage between the nature of the chemical bonding and the strength and sign of the magnetic coupling (1950). His picture of superexchange also explained the surprisingly large local molecular fields that Weiss had found in insulating materials (Section 2).

The importance of these data and the lucidity of what are they evidence remain undiminished to this day. The present author contests that there are few datasets, that were the *first in their field*, that we still continue to show so regularly to university students and at international conferences. These data also laid a clear challenge to crystallography – the space groups of Fedorov-Schoenflies were incapable of describing the symmetry of such an antiparallel arrangement of magnetic moments and how symmetry changed during an ordering transition. The symmetry descriptions used in crystallography and physics needed to be extended and generalised.

5 Magnetic moments as axial vectors

Before we embark on attempting to describe and predict the possible symmetries of a magnetic structure, that of an atomic magnetic moment must be defined. It was Pierre Curie that first did this, establishing that a magnetic field due to a current loop has the symmetry group of a rotating cylinder (1894). It therefore has a symmetry plane transverse to the axis of the current and a centre of symmetry. This symmetry-type is different to that of an ordinary (polar) vector as this has an infinite number of mirror planes longitudinal to its axis, but no transverse mirror plane, and contains the centre of symmetry. Woldemar Voigt coined the term 'axial vector' to describe this type of vector in his 1910 comprehensive book on crystallographic symmetry (1910).

In symmetry calculations, rather than consider the transformation of a current loop directly, it is convenient to write the transformation of an axial vector by a rotational symmetry operation, $R$, as $\boldsymbol{S'} = \det(\boldsymbol{R}).\boldsymbol{R}.\boldsymbol{S}$. For a polar vector the corresponding relationship is $\boldsymbol{S'} = \boldsymbol{R}.\boldsymbol{S}$. The difference between them, the determinant term, $\det(\boldsymbol{R})$, is +1 for proper rotations and -1 for improper rotations, mirror planes, and inversion. Its effect is to reverse the moment direction if the rotation involves inversion of the coordinate system (Figure 6).

As well as this classical construction of an axial vector to describe a magnetic field, it is sometimes helpful to consider an alternative definition using a vector $\boldsymbol{p}$ created by the cross product of two polar vectors, ***a*** and ***b***. This leads to the creation of a right-handed axis system:

$$\boldsymbol{p} = \boldsymbol{a} \times \boldsymbol{b} \qquad (5)$$

This cross product, *i.e.* the resultant vector ***p***, transforms under proper and improper rotation operations as an axial vector. For a spin-free quantum system, the axial properties of the magnetic moment then come from the transformation properties of the coordinate, or axis, system (Wigner, 1931).

However they are considered, researchers are encouraged to keep mind of Curie's note that an arrow does not depict properly the symmetry of a magnetic moment (as it has no transverse mirror plane or inversion symmetry). Emphasising the essential difference between magnetic moments and electric dipoles, he suggested that this could be done better as in Figure 7b, where a line is used to define an axis and magnitude of a magnetic field and the circulatory arrow indicates the sense of rotation.

6 Anti-symmetry and a generalisation of symmetry: black and white groups

The symmetry elements of the familiar point groups or crystallographic space groups operate on position coordinates. In 1930 the mathematician Heinrich Heech worked to extend this geometric aspect of group theory by introducing a new two-valued quality and a symmetry operation that could change it (Heesch, 1930). The difference in this 4th quality was termed *antisymmetry* and Heesch defined a new type of symmetry operation that could change it. This antisymmetry operation could be combined with the symmetry operations of the point groups, thereby extending the 32 point groups of 3-dimensional space to become the 122 point groups of a new 4-dimensional hyperspace.

Heesch developed the idea of antisymmetry as a mathematical construct to generalise the classical groups into higher dimensions. It was an abstract work that appeared too far from crystallography to find popular application. It would, however, reappear later in the work of Aleksei Shubnikov and Alexandr Zamorzaev. Shubnikov worked across the fields of crystal growth, the theory of symmetry, crystal structures, and the physical properties of crystals. Independently of the work of Heesch, Shubnikov developed antisymmetry as a tool to understand the spatial anisotropy of a crystal's physical properties, following in the direction laid by Curie (Shubnikov, 1951).

By allowing the antisymmetry operation to have arbitrary physical meaning, he developed an entirely new way of tackling a broad range of problems. His re-derivation of the 4-dimensional point groups (32 single colour groups, 32 grey (neutral) groups — those where every symmetry operation occurs both with and without antisymmetry, and 58 black and white groups — those where some antisymmetry was combined with some symmetry operations, Figure 8) yielded the 122 generalised point groups, now classified as *colourless* (normal), *grey*, and *black and white*. He accompanied them by geometric figures that made clear their relevance to crystallography and so facilitated their widespread dissemination.

Under the guidance of the geometer Aleksandr Alexandrov, Zamorzaev would later expand upon Shubnikov's ideas of antisymmetry in his thesis *Generalization of Fedorov groups* (Zamorzaev, 1953; Zamorzaev, 1957), where he derived the 4 types of colour space groups (monochromatic, grey, monochromatic Bravais lattice and dichromatic point group, and dichromatic Bravais lattice – Figure 9). This work extended the space groups from the 230 normal groups to 1651. Later, alternative derivations and analyses were made, such as those by Belov (Belov *et al*, 1955; Belov and Tarkhova, 1956; Belov *et al*, 1957), Koptsik (1966), and Indenbom (1959) and Niggli (1959). The nomenclature of these 4-dimensional point groups and space groups suffers from history. By some they were referred to as 'Shubnikov groups', to honour his reinvention of antisymmetry, but this nomenclature fails to acknowledge Heesch's earlier work and Zamorzaev's initial derivation. Similarly, use of 'magnetic space groups' fails to relate that these 4-dimensional space groups were developed as a general construction with applications in different aspects of crystallography and physical properties. The arbitrary two-valued state could, for example, relate to the sign of a charge or the direction of a displacement vector or an electric dipole. It was simply in magnetism that they found immediate popularity as the magnetic structure proposed for MnO could now be described by defining antisymmetry as an operation that reverses atomic moments or spins (Tavger and Zaitsev, 1956), thereby solving the challenge of describing magnetic symmetries, at least in part.

primitive tetragonal unit cell (Izyumov, 1980).

7 Construction of the black and white space groups

As we have seen, the black and white groups involve a simple antisymmetry operation that changes the colour value. A one-to-one correspondence allows them to be derived simply from the one-dimensional irreducible representations (IRs) of the parent symmetry groups by selecting possible subgroups of index 2, *i.e.* the antisymmetry operation, $\theta$, is added to the symmetry operations with character '-1' (Indenbom, 1959; Niggli, 1959; Bertaut, 1968; Sivardière, 1969). Commonly, the operation is written as being primed. This is exemplified for the two dimensional point group $4mm$, the character table of which is given in TABLE I.

TABLE I: Character table of the IRs for the point group $4mm$ (Cracknell, 1969).

|       | $E$ | $C_{2z}$ | $C_{4z}^+, C_{4z}^-$ | $\sigma_z, \sigma_V$ | $\sigma_{da}, \sigma_{db}$ |
|-------|-----|----------|----------------------|----------------------|----------------------------|
| $A_1$ | 1   | 1        | 1                    | 1                    | 1                          |
| $A_2$ | 1   | 1        | 1                    | -1                   | -1                         |
| $B_1$ | 1   | 1        | -1                   | 1                    | -1                         |
| $B_2$ | 1   | 1        | -1                   | -1                   | 1                          |
| $E$   | 2   | -2       | 0                    | 0                    | 0                          |

The identity IR ($A_1$), which has unit character for all the symmetry operations, corresponds to the colourless point group:

$$4mm = E, C_{4z}^+, C_{4z}^-, C_{2z}, \sigma_x, \sigma_y, \sigma_{da}, \sigma_{db}$$

Priming all of the operations with character '-1' in the other 1-dimensional IRs gives the black and white point groups:

$$(B_1)4'mm' = E, \theta C_{4z}^+, \theta C_{4z}^-, C_{2z}, \sigma_x, \sigma_y, \theta\sigma_{da}, \theta\sigma_{db}$$

and

$$(A_2)4m'm' = E, C_{4z}^+, C_{4z}^-, C_{2z}, \theta\sigma_x, \theta\sigma_y, \theta\sigma_{da}, \theta\sigma_{db}$$

That corresponding to $B_2$ is related to $B_1$ by axis rotation. Geometric depictions of $4'mm'$ and $4m'm'$ are shown in Figure 8.

.

8 Generalisation to the coloured space groups

This relationship between one-dimensional IRs and coloured symmetry groups extends further than is often realised. While complex IRs no longer correspond to the black and white groups as the character of the IRs can involve complex numbers rather than only ±1, Indenbom (1959) and Niggli (1959) suggested independently that these could be used to derive the multicolour (polychromatic) symmetry groups that had been conceived by Belov (Belov *et al*, 1955; Belov and Tarkhova, 1956). (Figure 10). In these groups repeated operation of the colour-changing operation causes a cycling through a series of $p$-colours ($p > 2$), thereby further generalising the anti-symmetry operation of the black and white groups.

Another generalisation, that we will return to in a following discussion of magnetic superspace groups, can be introduced by changing how the symmetry operations act. In creating the black and white groups, the new symmetry operation of antisymmetry was added to some of the normal symmetry operations to form a compound primed operation that acts on both the atomic position and the two state physical characteristic. This type of structure was termed the *P*-type colour groups, cyclic symmetry groups, or quasisymmetry groups (Shubnikov and Koptsik, 1972; Zamorzaev, 1988; Krishnamurty, 1978) and they involve a group of colour permutations. Another possibility is that the symmetry operations are separated into ones that act on the atomic positions (geometric space) and others that affect the physical characteristic (non-geometric oeprations). These were termed spin groups or spin-space groups depending on the authors (Belov and Tarkhova, 1956; Naish, 1962; Kitz, 1965; Brinkman and Elliot, 1966; Opechowski and Dreyfus, 1971; Landau and Lifshitz, 1957).

When dealing with magnetic structures there is physical significance in the applicability of *P* and *Q*- type groups: they correspond to different situations of spin-orbit coupling (Brinkman and Elliot, 1966). The *Q*-type group is less restricted as the group acting on the atomic spins can contain operations that are not in the group applied to the atomic positions – it is suited to isotropic spins. The *Q*-type groups then correspond to situations when the spin-orbit coupling and any spin anisotropy terms are weak, such as for the isotropic 3-dimensional group of spin rotations O(3). The operations that act upon the spins in *P*-type groups are more restricted and so these groups are suited to situations with strong spin-anisotropy as then the spins can only point along limited directions.

9 Antisymmetry and time- reversal

The antisymmetry of Heesch and Shubnikov is a general property that had arbitrary physical significance. It was Lev Landau and Evgeny Lifschitz that cast it into a particular form when they described it as the operation of 'time reversal' in a discussion on magnetic symmetry (1951) and this led to a reinvestigation of the two-colour groups as magnetic groups (Tavger and Zaitsev, 1956). This image where antisymmetry is linked with time reversal is one borrowed from classical physics where an atomic magnetic moment is considered as arising from a circulating electric current, just as in Langevin's models for diamagnetism and paramagnetism. Reversing time would correspond to a reversal in the direction of electric current and so of the atomic moment. By introducing time reversal symmetry in this way, Landau and Lifshitz also made a clear connection to the theory of symmetry-breaking phase transitions as magnetic ordering then breaks time reversal symmetry. For example, when the atomic-level magnetic moments of a collinear ferromagnet point along the *c*-axis, the local $c$ and $-c$ directions of the crystal structure are no longer equivalent.

The present author councils against using this definition of antisymmetry, as the operation of time reversal within this classical picture it does not survive generalisation to the multicolour groups. Its also hinders understanding of the role that time reversal plays within quantum mechanics, where it can be regarded as the operation of complex conjugation rather than to reverse magnetic moments.

Figures

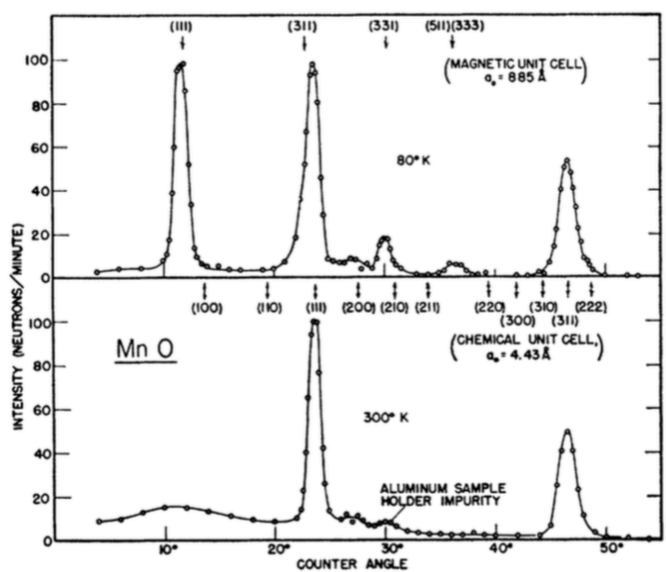

FIGURE 1. The first magnetic neutron diffraction data, taken from MnO at 80 K, reveal that four new diffraction peaks appear upon cooling below the magnetic ordering transition at 120 K (Shull and Smart, 1949) .

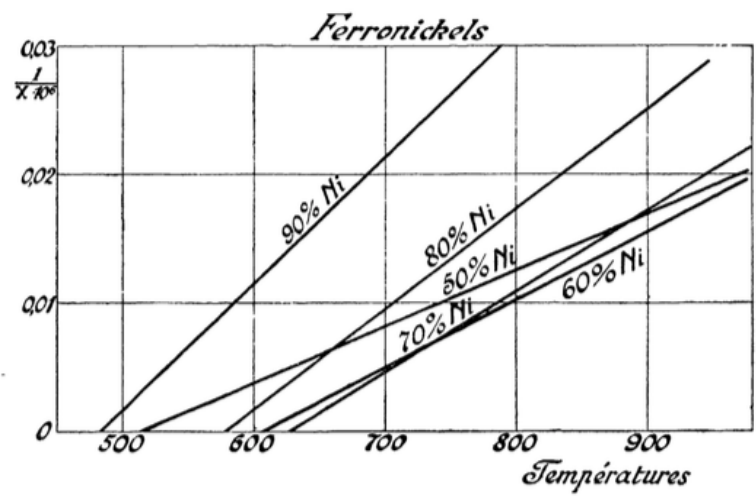

FIGURE 2. The commonly observed linear decrease in the inverse susceptibility upon cooling became known as the *Curie law*. This dependence is exemplified by data taken from a variety of iron/nickel alloys (Weiss and Foex, 1911).

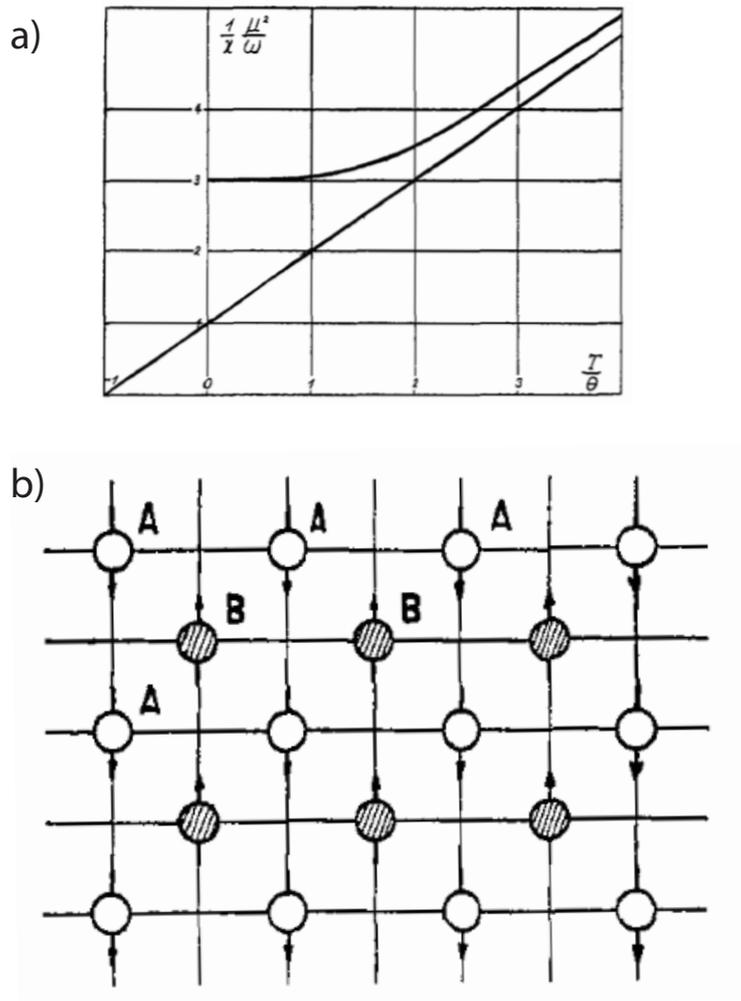

FIGURE 3. (a), Néel proposed that the inverse susceptibility of a magnet with negative local molecular field was a constant below an ordering transition to a magnetic structure. (b) The magnetic structure was made up of two-sublattices of antiparallel spins, labelled *A* and *B* (Néel , 1936).

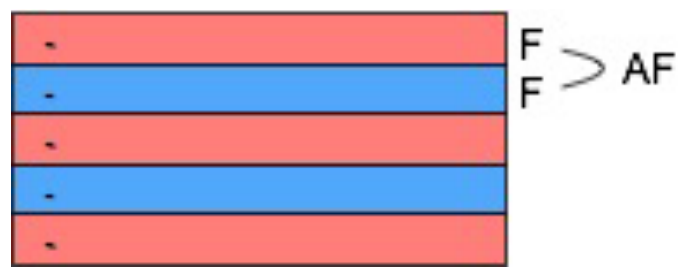

FIGURE 4. Landau modelled the layered structures of various transition metal halides as strongly interacting layers (ferromagnetically coupled) that were negatively coupled by a much weaker energy (Landau, 1933).

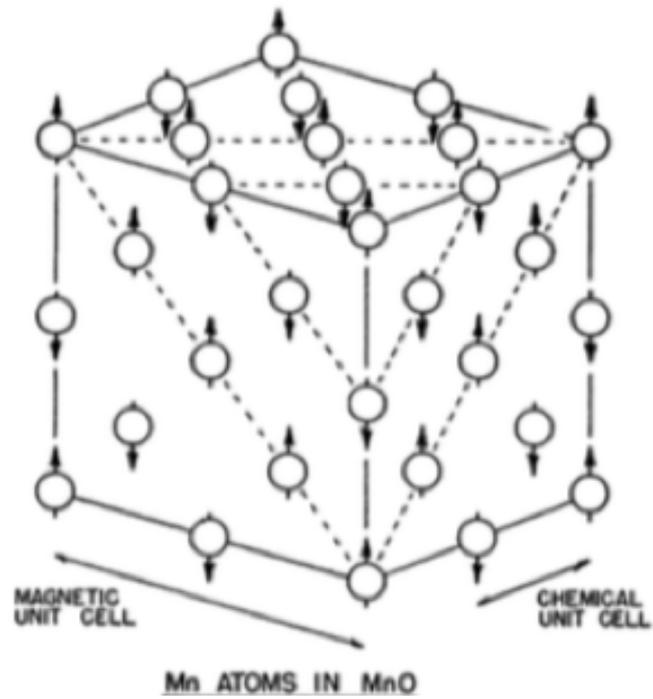

FIGURE 5. The antiferromagnetic structure of MnO has a unit cell with a lattice parameter twice that of the chemical unit cell. Neighbouring moments form an antiparallel series, following the predictions of Néel. Magnetic interactions between the Mn ions occur via intermediate oxygen atoms rather then direct exchange, following the mechanism of superexchange [14].

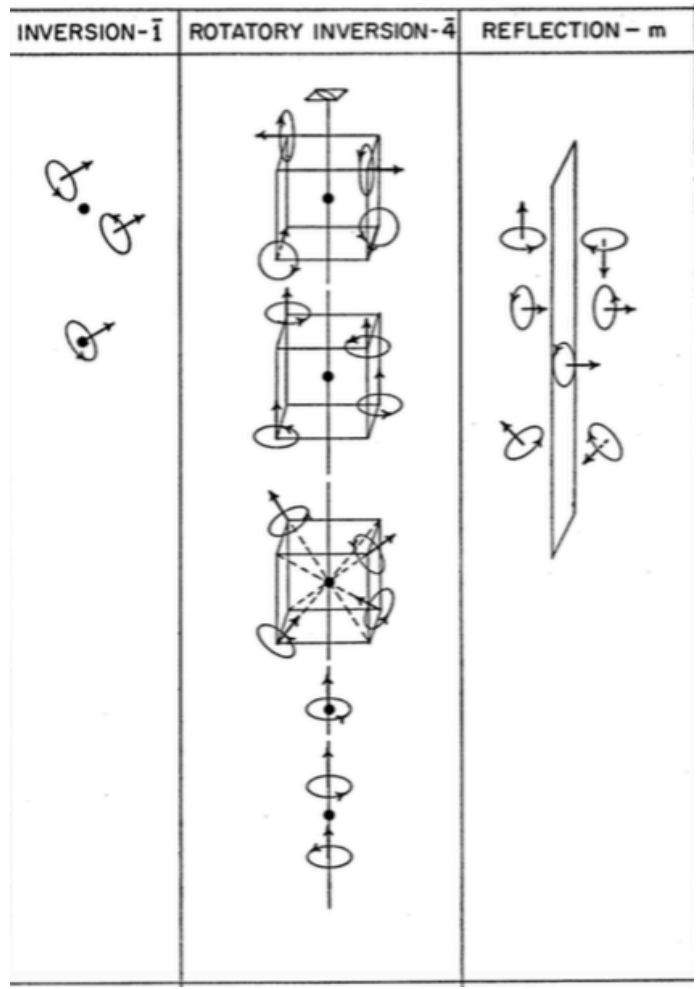

FIGURE 6. A representation of how axial vectors are transformed by various symmetry operations. Adapted from (Donnay *et al*, 1958).

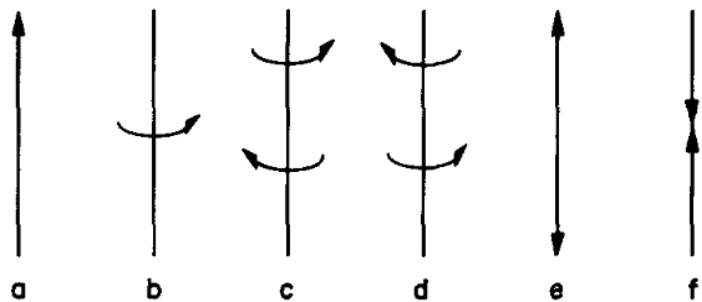

FIGURE 7. Graphical depictions of directional quantities: (a) polar vector, *e.g.* strength of an electric field or an atomic displacement; (b) axial vector, *e.g.* strength of a magnetic field; (c, d) axial tensor, *e.g.* the magnitude of the left and right specific rotation of the polarisation plane; (e, f) polar tensor, *e.g.* tensile and compressive stresses (Shubnikov, 1988).

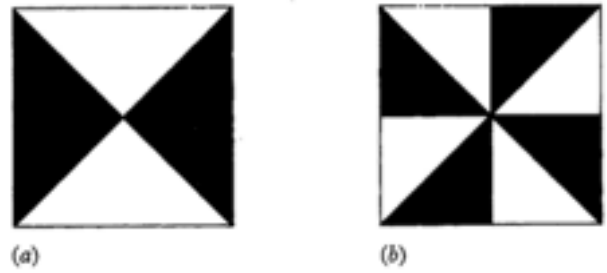

FIGURE 8. Black and white squares to illustrate the back and white point groups (a) $4'mm'$ and (b) $4m'm'$ (Cracknell, 1969).

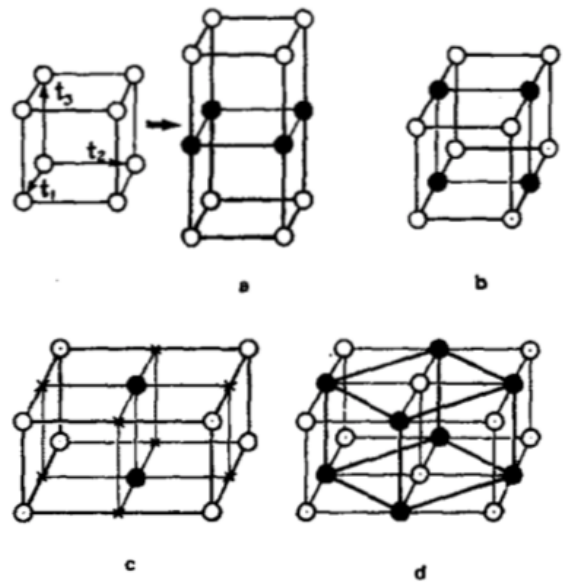

FIGURE 9. The black and white lattices, or dichromatic Bravais lattices, created from a

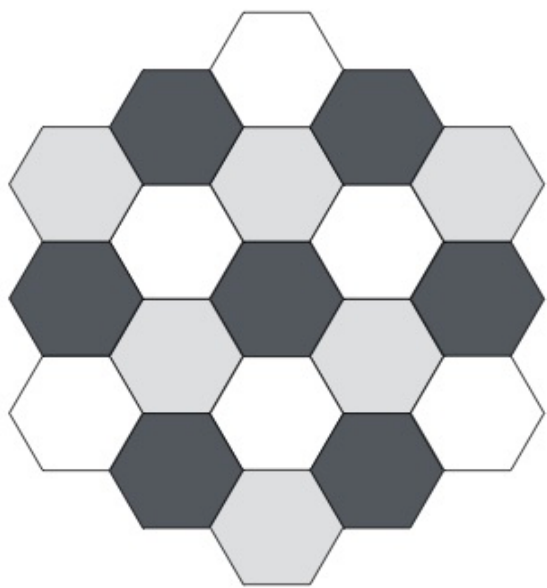

FIGURE 10. Belov's 3-colored pattern of the tiling of the plane by regular hexagons [42].

The tiling pattern is said to be 'perfect' as every operation in $G$ is associated with a unique colour permutation (Senechal, 1988).